\documentclass[aps,floatfix,groupedaddress,longbibliography,amsmath,amssymb,pre,twocolumn]{revtex4-1}
\usepackage[utf8]{inputenc}
\usepackage{graphicx}

\newcommand{\step}{\operatorname{step}}

\newcommand{\erf}{\operatorname{erf}}
\newcommand{\dry}{\operatorname{dry}}
\newcommand{\wet}{\operatorname{wet}}

\begin{document}

\title{Optimizing coating performance for diffusion under cyclic moisture exposure}
\author{Aaron J.~Feickert}
\email{\texttt{aaron.feickert@ndsu.edu}}
\author{Alexander J.~Wagner}
\email{\texttt{alexander.wagner@ndsu.edu} (corresponding author)}
\affiliation{Department of Physics, North Dakota State University, NDSU Dept 2755, PO Box 6050, Fargo ND 58108-6050, USA}

\date{\today}

\begin{abstract}
Fickian diffusion is often used to model moisture transport through barrier coatings, where the goal is to protect an underlying substrate from the onset of corrosion caused by buildup of water or other aggressive species. Such coatings are often exposed to cyclic moisture, either in laboratory testing or in service due to natural environmental fluctuations. In this paper, we use lattice Boltzmann numerical techniques to investigate the effects of reservoir cycling on moisture propagation and concentration at the substrate where corrosion onset occurs. We examine both the simple case of constant diffusivity, representing idealized Fickian diffusion, and diffusivity that depends on concentration via either a step or linear function, representing polymer network swelling. The use of a coating subject to swelling is shown to lead to highly variable equilibrium behavior. We show that the nature of the functional diffusivity has large effects on water concentration at the substrate, and has implications for material design and analysis to avoid corrosion.
\end{abstract}

\maketitle

\section{Introduction}
Polymeric coating systems are used widely for infrastructure and material protection. Aggressive environmental factors like moisture, salt, chemical cleaners and de-icers, radiation, and mechanical stress are common in a variety of field conditions where long coating life and substrate structural integrity are essential. The variety of environments to which a coated system must be exposed implies that the particular factors impacting a system will be different and depend on location, season, and purpose. Crosslinked polymer network structures are chosen for use as barrier coatings because of their net-like structure, helping to decrease effective diffusivity and swelling and slow moisture ingress.

Moisture is a key component of overall coating failure in field applications. Overall uptake enhances photooxidation and hastens degradation processes that decrease service life \cite{Hardcastle2008}, both alone and in the presence of ultraviolet radiation from the sun \cite{Baukh20131197}. Transport of moisture, and especially the time scales at which it occurs, has broad effects on the thermal and mechanical properties of coating systems over time \cite{Shi2010}, as well as on the adhesion of a coating to other layers or a substrate \cite{Kamisho2014}. Moisture permeation has also been linked to the presence of pores \cite{AIC:AIC690480104} and pathways \cite{Taylor2012169} within a coating that are often the result of the statistics and chemistry of the crosslinking process \cite{Kroll201582} both before and after gelation \cite{Kroll2017154,Kroll2017113}, and of cavitation \cite{Zee201555}.

Cyclic weathering testing is a common and accepted part of the development and analysis of barrier coatings intended for substrate protection. This testing frequently includes multiple variables, including water or salt spray, radiation, temperature, and humidity. Despite the natural diurnal variance of these factors in nature, the accelerated nature of testing means cycles are much shorter than day or night periods, often as short as a few hours. Since corrosion is linked to moisture reaching the substrate in sufficient quantities, understanding the nature of substrate exposure and coating saturation over time is essential to better weathering testing. However, some tests do not permit most barrier coatings to reach saturation within a cycle due to the short time scales used \cite{Hardcastle2008}. In general, cyclic testing does not provide a complete picture of the failure modes or service life of a system under test \cite{doi:10.1021/bk-2002-0805.ch001}.

Modeling the kinetics of moisture transport through a coating or stack permits a fine-grained approach to examining water content throughout the system over time, which is difficult to achieve in the laboratory \cite{vanderWel19991}. Experimental methods like electrochemical impedance spectroscopy are well suited to examine bulk properties like diffusivity and uptake during cycled exposure \cite{Hinderliter2008}, but have the disadvantage of requiring the constant presence of an electrolyte for measurement. However, it is straightforward to gain information about the diffusion process. Fickian diffusion has been shown to provide a good basis for water transport through a variety of barrier and underlying base coatings with different chemistries \cite{Kamisho2014,Hinderliter2008,Hardcastle2008,APP:APP15}. Other work has examined bilayer systems where an instaneous reservoir model was used for a hydrophilic base coat \cite{Baukh20123304}. Multilayer systems provide additional modeling challenges, since additives like pigments can lead to a variety of unexpected changes in diffusivity \cite{Perera2004247} and boundary conditions must be carefully considered \cite{de1998monitoring}. However, an assumption of generally Fickian behavior provides good matching to experiment under a variety of setups.

In this paper, we apply a lattice Boltzmann numerical method to the problem of a single-layer coating exposed to a variable reservoir and perfectly adhered to an impermeable substrate. We use Fickian diffusion to examine the kinetics of wetting and drying over time, and provide a simple scaling argument that permits us to extend the model to a variety of experimental setups and parameters. We introduce reservoir cycling to determine the effects on substrate wetting and eventual oscillatory steady-state behavior. Despite the lack of a unified theory of concentration dependence \cite{vanderWel19991}, several cases are considered for this dependence of the diffusivity: constant diffusivity, a step function induced by instantaneous network swelling, and a linear dependence induced by gradual swelling. Variable cycle time ratios are considered throughout. We conclude by proposing material properties for coatings that may be particularly beneficial to limit corrosion.

\section{Theory}
In this section, we show the derivation of an analytical solution to the diffusion equation with correct boundary conditions. We then outline the numerical lattice Boltzmann method used for subsequent results.

\subsection{Analytical}
We wish to model an idealized coating that is adhered to an impermeable substrate and exposed to either a moisture reservoir or to air. This models the common use case where a coating might be applied to an airframe or automobile, and thereafter be exposed to the elements. Additionally, it models the cycled environmental exposure that most coating systems undergo in test chambers during laboratory testing, where the coating might experience repeated cycles of water misting followed by dry exposure to ultraviolet radiation. Parameters like temperature, timing, and radiation intensity of each cycle segment vary depending on the testing protocol; perhaps unsurprisingly, most protocols align cycle times with the workday (\textit{e.g.} 8, 12, or 24 hours) to more easily accommodate predictable personnel availability.

Mathematically, the coating panel is modeled as a simple one-dimensional system, where we discount edge effects and assume that moisture only reaches the substrate by diffusing through the bulk of the overlying coating via Fickian diffusion, where the concentration $\rho$ follows the diffusion equation
\begin{equation}
\partial_t \rho(x,t) = \nabla\cdot(D(\rho)\nabla\rho(x,t))
\end{equation}
for diffusivity $D(\rho)$. Since test panels represent a framed wide area in the test chamber relative to thickness, panel edge effects are negligible, so the use of a one-dimensional model is reasonable and matches the masking effect of the frame. In a first approximation, the diffusivity $D(\rho)$ may be considered a constant that does not vary in space, time, or concentration:
\begin{equation}
\partial_t \rho(x,t) = D\nabla^2\rho(x,t)
\end{equation}
As we discuss later, however, polymer network swelling under the influence of a solvent may lead to concentration-dependent diffusivity, where such a simplification to the diffusion equation is no longer possible.

The model one-dimensional coating system extends in the range $0 \leq x \leq L$ for some coating thickness $L$. At $x=0$ is an infinite reservoir whose concentration may vary with time. At $x=L$ is an impermeable substrate where the concentration gradient is always zero. No particular assumptions are made regarding the adhesion of the coating to the substrate, and the coating is assumed to have uniform density and saturation capacity.

To derive an analytical solution giving the moisture concentration at any point over time, we first consider a simpler model, consisting of an infinite system with an initial step function concentration of $2\rho$ for $x<0$, and a concentration of zero for $x>0$. The concentration evolution over time $\rho_{\step}$ is given by the well-known error function expression \cite{crank1979mathematics}
\begin{equation}
\rho_{\step}(x,t) = \rho_0\left[1 - \erf\left(\frac{x}{\sqrt{4Dt}}\right)\right].
\end{equation}

\begin{figure}
\centering
\includegraphics[width=\columnwidth,clip=true]{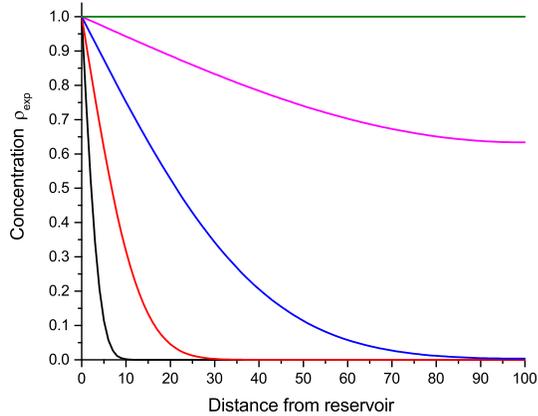}
\caption{Example of concentration $\rho_{\operatorname{exp}}$ during constant exposure, going from mostly dry (lowest curve) to completely saturated (highest curve) over time}
\label{fig:exposure}
\end{figure}

The presence of a fixed point $\rho_{\step}(x=0,t>0) = \rho_0$ means we may account for an infinite reservoir of fixed concentration $\rho_0$ simply by restricting our solution to the range $x>0$. Finally, we must account for the impermeable substrate. To incorporate this boundary, we add an image source symmetric to the $x=L$ substrate boundary. Since this image source will eventually diffuse back to the reservoir at $x=0$, we must subtract an image sink symmetric to the reservoir. Continuing this image process infinitely, we arrive at the concentration $\rho_{\operatorname{exp}}$ that models constant moisture exposure and includes both reservoir and substrate \cite{2017arXiv170305795S}:
\begin{multline}
\rho_{\operatorname{exp}}(x,t) = \rho_0\sum_{i=0}^\infty (-1)^i\biggl[ 2 + \erf\left(\frac{x-2(i+1)L}{\sqrt{4Dt}}\right) \\
- \erf\left(\frac{x+2iL}{\sqrt{4Dt}}\right) \biggr]
\label{eqn:rhoexp}
\end{multline}
Figure \ref{fig:exposure} shows an example of this concentration over time as the system goes from dry (lowest curve) to saturated (highest curve) over time.

\begin{figure}
\centering
\includegraphics[width=\columnwidth,clip=true]{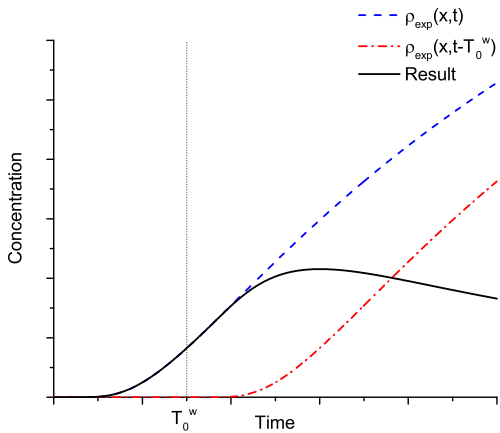}
\caption{Construction of single-cycle concentration at lattice position $x=L/2$, showing constant-exposure $\rho_{\operatorname{exp}}(x,t)$ (dashed blue), time-shifted constant-exposure $\rho_{\operatorname{exp}}(x,t-T^w_0)$ (dot-dashed red), and resulting difference (solid black)}
\label{fig:rho}
\end{figure}

This solution assumes the reservoir remains at fixed concentration, modeling the case where the coating undergoes constant exposure and is eventually saturated to concentration $\rho_{\operatorname{exp}}(0 \leq x \leq L,t \to \infty) = \rho_0$. Suppose that instead, the reservoir is removed at some time $T^w_0$ and the coating is allowed to dry. We may express the concentration over time by modifying the result of Equation \ref{eqn:rhoexp}. If we let $\rho_{\operatorname{exp}}(x,t<0) = 0$, then the concentration at any time (before or after the reservoir is removed) may be expressed by the superposition
\begin{equation}
\rho_{\operatorname{exp}}(x,t) - \rho_{\operatorname{exp}}(x,t-T^w_0).
\end{equation}
Figure \ref{fig:rho} shows the construction of this expression at a representative position over time (arbitrary units).

In the general case where the wet and dry timings are arbitrary and not necessarily constant between cycles, let $T^w_0,T^w_1,\ldots$ be the lengths of each wet period, and $T^d_0,T^d_1,\ldots$ the lengths of each dry period. That is, the reservoir turns on at times $t = 0, (T^w_0+T^d_0), (T^w_0 + T^d_0 + T^w_1 + T^d_1),\ldots$ according to the given timings. The solution then becomes
\begin{multline}
\rho_{\operatorname{cycle}}(x,t) = \rho_{\operatorname{exp}}(x,t) \\
+ \sum_{i=1}^\infty\biggl[ \rho_{\operatorname{exp}}(x,t-t^w_i) - \rho_{\operatorname{exp}}(x,t-t^d_i) \biggr]
\label{eqn:rhocycle}
\end{multline}
where we define the partial sums $t^w_i = \sum_{j=0}^i (T^w_j+T^d_j)$ and $t^d_i = t^w_i-T^d_i$.

\subsection{Lattice Boltzmann}
The analytical theory gives the precise concentration throughout the coating at any time for the case of constant diffusivity. However, it does not easily accommodate the scenario where diffusivity within a coating layer varies with concentration due to polymer network swelling under the influence of a solvent.

To incorporate this important phenomenon, we employ a diffusive lattice Boltzmann method previously introduced \cite{2017arXiv170305795S}. Such lattice approaches are used to model a variety of phenomena, such as hydrodynamics \cite{PhysRevLett.56.1505,qian1992lattice,PhysRevLett.75.830}, diffusion \cite{shan,wolfdiffusion}, and electrostatics \cite{electrostatics}. While we defer the interested reader to the paper \cite{2017arXiv170305795S} for a more thorough discussion of lattice Boltzmann diffusion, we present a brief outline here. The method models the system as a one-dimensional lattice, where particle densities $f_i$ move between lattice sites under the influence of collisions. By using the so-called BGK collision operator, the density evolution is given \cite{qian1992lattice} by the lattice Boltzmann equation
\begin{equation}
f_i(x+v_i,t+1) = f_i(x,t) + \frac{1}{\tau}\left[f_i^0(x,t) - f_i(x,t)\right]
\end{equation}
for local equilibrium $f_i^0$ and collision relaxation parameter $\tau$, where $v_i$ represents particle density velocity.

Since we are considering only a one-dimensional case, we use the simplest lattice Boltzmann model with the velocity set $\{v_i\} = \{0,1,-1\}$.  At any lattice site $x$ at time $t$, the moisture concentration is given by $\rho(x,t) = f_{-1}(x,t) + f_0(x,t) + f_1(x,t)$. This lattice with given velocities is known in the literature as a D1Q3 scheme, where we express the local equilibrium distribution as $f_i^0 = \rho w_i$ for weights associated to each velocity. We use the weights
\begin{eqnarray}
w_0 &=& 1 - \theta \\
w_1 &=& \frac{\theta}{2} \\
w_{-1} &=& \frac{\theta}{2}
\end{eqnarray}
to recover the necessary distribution moments, where $\theta$ is a system parameter. This choice of equilibrium distribution leads, via a Taylor approximation to second order \cite{PhysRevE.94.033302}, to the lattice diffusion equation
\begin{equation}
\partial_t \rho = \nabla\left(\tau - \frac{1}{2}\right)\nabla(\rho\theta)
\end{equation}
where the diffusion constant is given, for a spatial region of constant $\tau$ and $\theta$, as
\begin{equation}
D = \left( \tau - \frac{1}{2} \right)\theta.
\label{eqn:D}
\end{equation}

\subsection{Scaling and dimensional analysis}
Introduce length, time, and concentration scales $x'$,$t'$, and $\rho'$ such that
\begin{eqnarray}
t &=& Tt' \\
x &=& Xx' \\
\rho &=& \rho_0\rho'
\end{eqnarray}
and $0 \leq \{t',x',\rho'\} \leq 1$. Dimensional analysis gives that the value
\begin{equation}
F \equiv \frac{TD}{X^2}
\end{equation}
is a dimensionless quantity, and is a constant for any particular experimental setup. For simulations where one wishes to vary $\tau$ or $\theta$, for example to control variable diffusivity, simulation times must be scaled by the value $T$ for result comparison.

Below we discuss scaling values representative of a sample environment, and control for this in subsequent lattice Boltzmann numerical simulations. We use these conditions to establish the dimensionless quantity $F$, and scale other parameters accordingly.

\subsection{Boundary conditions}
The substrate underlying the coating in our model is assumed to be impermeable, so we use a reflection condition for lattice density evolution there. After each collision, any $f_1$ outflow attempting to pass through the substrate at $x = L$ is reflected back as $f_{-1}$ inflow to the adjacent lattice site. Since this means the boundary is effectively located half a lattice space outward, the corresponding analytical solution is slightly modified to assume the system is correspondingly longer.

Earlier work \cite{2017arXiv170305795S} found that the choice of reservoir boundary condition is critical for determining the overall numerical error, especially near the boundary and at early times. In particular, manually setting the reservoir boundary lattice densities $\{f_i\}$ to the equilibrium distribution value at each timestep resulted in substantially larger errors. One proposed solution, that of a periodic embedding of the system where such boundary conditions are replaced by symmetric initial conditions, removes such error almost completely.

Extending this embedding to time-dependent boundary conditions (such as reservoir cycling) is possible, but we found an even simpler way of defining an inflow boundary condition in the finite lattice that retains the accuracy of the periodic embedding and keeps the system size smaller for computational efficiency. The reservoir boundary condition is modified to reflect $f_{-1}$ outflow densities about the current reservoir concentration to $f_1$ inflow, instead of using the equilibrium distribution value:
\begin{equation}
f_1(x=0) = 2\rho_0\frac{\theta}{2} - f_{-1}(x=0)
\end{equation}
It was found that this boundary condition retains the same numerical accuracy as a periodic embedding.

Due to parameter scaling, a single coating layer whose diffusivity is constant is trivial to model, since any choice of parameters $\tau$ and $\theta$ is analytically equivalent. However, previous work \cite{2017arXiv170305795S} showed that straying from $\tau = 1.0$ to higher values introduces rapidly increasing error, making it an obvious choice. For this work, we use $\tau = 1.0$ and $\theta = 0.5$ when constant diffusivity is assumed, and scale to values $\tau \leq 1.0$ when working with non-constant diffusivity.

Many barrier coatings exhibit swelling, where the presence of solvent causes the crosslinked polymer network to expand and permit faster moisture ingress and egress. Conversely, one might imagine a metamaterial designed such that solvent might close any pore structures in the backbone and reduce effective diffusivity. Analytically, either case corresponds to defining $D = D(\rho)$ according to some functional form. There is no known unified model for this swelling behavior from first principles \cite{vanderWel19991}, but two common models of polymer swelling are a step function front and linear diffusivity \cite{Pathania2017149}. In the step function model, the coating is assumed to be in a dry state when at low moisture concentration until a critical concentration is reached, above which it is in a wet state and at a different diffusivity. This means $D(\rho)$ takes the form of a step function. The other common functional form is that of a line, where diffusivity varies monotonically with concentration. We consider both cases in the subsequent analysis.

\section{Results and discussion}
\subsection{Constant diffusivity}
A typical single-layer barrier coating might be applied with thickness 50 $\mu$m to a test panel, have a diffusion constant in water of $D \sim 10^{-14}$ m$^2/$s, and be exposed to moisture in an environmental chamber for four hours (14400 s). This corresponds to a dimensionless constant $F = 5.76 \times 10^{-2}$. For simulations, we choose a lattice of size $L = 100$ lattice points, reservoir concentration $\rho_0 = 1$, relaxation time $\tau = 1.0$, and $\theta = 0.5$. This means that four hours of moisture exposure in the macroscopic system corresponds to $T \approx 3500$ timesteps in the lattice Boltzmann system. It should be noted that diffusion constants for water through barrier coatings have large variance, with values that may range as low as $10^{-16}$ m$^2/$s; due to the ease of scaling the problem, this does not pose a significant issue for our analysis.

\begin{figure}
\centering
\includegraphics[width=\columnwidth,clip=true]{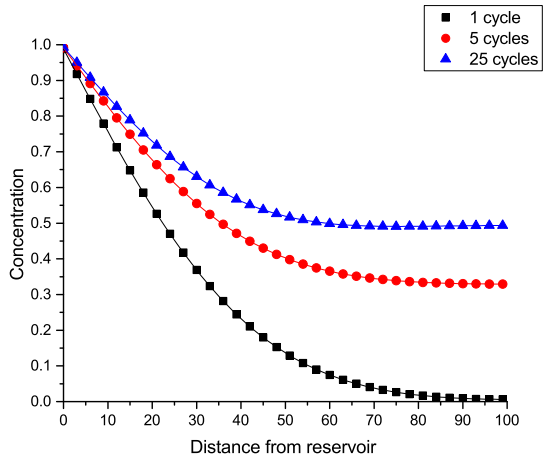}
\caption{Numerical (symbols) and theoretical (Equation \ref{eqn:rhocycle}) (lines) concentration at the end of wet cycles over simulation lattice}
\label{fig:wet-cycles}
\end{figure}

\begin{figure}
\centering
\includegraphics[width=\columnwidth,clip=true]{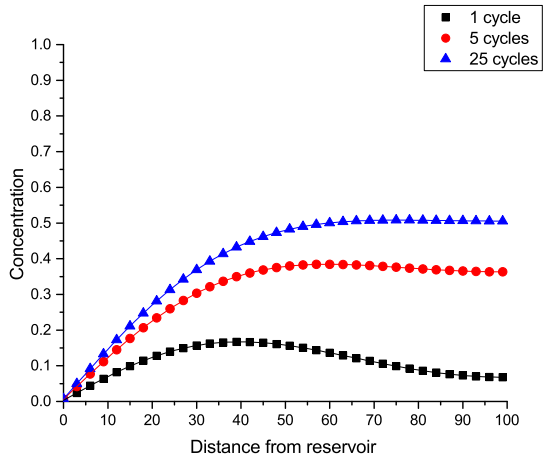}
\caption{Numerical (symbols) and analytical (lines) concentration at the end of dry cycles over simulation lattice}
\label{fig:dry-cycles}
\end{figure}

It has previously been shown \cite{2017arXiv170305795S} that this setup results in excellent numerical accuracy for moisture exposure over long times, typically on the order of $0.01\%$ of the predicted theory value, several orders of magnitude better than typical experimental measurements. To verify that cycling the reservoir, which introduces large concentration gradients near the reservoir, retains the desired numerical accuracy to theory, we run a lattice Boltzmann simulation with the given parameters, cycling the reservoir on and off every $T$ timesteps for a total cycle period of $2T$ timesteps. Snapshots of the lattice and corresponding theory from Equation \ref{eqn:rhocycle} are shown in Figure \ref{fig:wet-cycles} for the end of wet cycles (just before the reservoir turns off), and in Figure \ref{fig:dry-cycles} for the end of dry cycles (just before the reservoir turns on). Even at longer times when the coating concentration reaches full saturation, the numerical solution shows excellent matching to theory.

The onset of corrosion is commonly expected once the substrate becomes sufficiently wet. However, it remains an open question at what moisture content corrosion onset is expected. We examine the substrate concentration as it evolves toward a periodic late-time regime by fixing the total wet-and-dry cycle length at $2T = 7000$ timesteps (equivalent to eight hours in the corresponding macroscopic system) and varying the wet-to-dry cycle time ratio. In the periodic late-time regime, the inflow and outflow at any point in the lattice occur at the same rate. The concentration averaged over a period is the same throughout the lattice. Since the concentration oscillates about a value that is scaled by the relative time averages at the reservoir, we also know that this will be the same average value at the substrate. That is, for a wet-to-dry cycle ratio of 1:$R_t$ for some ratio $R_t$, the late-time average concentration value should be $1/(R_t+1)$.

\begin{figure}
\centering
\includegraphics[width=\columnwidth,clip=true]{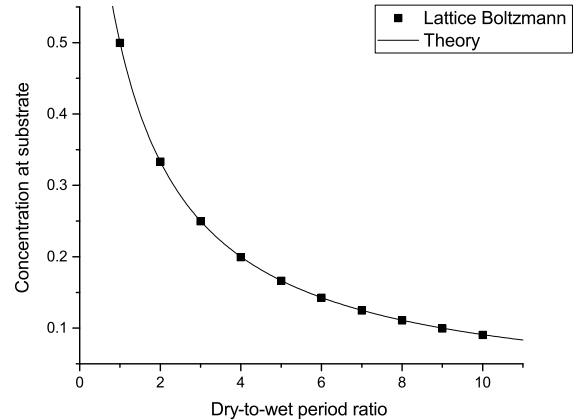}
\caption{Constant diffusivity steady-state average substrate concentration (symbols) with varying wet-to-dry cycle timings and total cycle length $2T = 7000$ timesteps, with comparison to $1/(R_t+1)$ theory (line)}
\label{fig:single-var-cycle}
\end{figure}

For simulations, we divide the time evolution into windows of length $10T$; we say the system has reached a steady state when the maximum and minimum values within a window are within $0.01$ of the values in the next window. For reference, with a 1:1 cycle time ratio, the system comes within $2\%$ of the asymptotic value after about $13$ full moisture cycles, corresponding in our equivalent macroscopic system to slightly over four days of cycled exposure. We then compute the average substrate concentration from that point. Results are shown in Figure \ref{fig:single-var-cycle}. Reassuringly, the final concentration scales precisely with the cycle time ratio.

\subsection{Variable diffusivity: step function}
Diffusion of water through some polymeric coatings may be modeled by an instantaneous change in diffusivity that occurs at a particular critical concentration $\rho_c$. The diffusivity $D(\rho)$ takes on a constant value $D_{\dry}$ below $\rho_c$, and another value $D_{\wet}$ above $\rho_c$. When examining a single-layer coating in this way, we therefore have a three-parameter step function:
\begin{equation}
D(\rho) = \left\{ \begin{array}{lr} D_{\dry}, & \rho < \rho_c \\ D_{\wet}, & \rho \geq \rho_c \end{array} \right.
\end{equation}
However, for our choice of dimensionless constant $F \sim 0.06$ in which oscillations at the substrate are small, we need only be concerned with diffusivity ratios due to scaling. It suffices to reduce to a two-parameter function where the saturated diffusivity is scaled
\begin{equation}
R \equiv \frac{D_{\wet}}{D_{\dry}}
\end{equation}
from the dry value. Most polymeric materials swell under the influence of solvent and permit faster moisture transport, so in this model such coatings would have $R>1$. We also present results for $R<1$, and discuss the implications in Section \ref{sec:material}.

\begin{figure}
\centering
\includegraphics[width=0.6\columnwidth,clip=true]{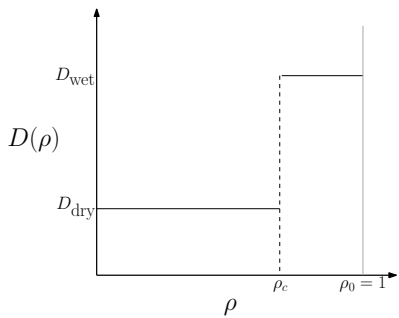}
\caption{Three-parameter step function $D(\rho)$, indicating two diffusivity values and critical concentration}
\label{fig:step}
\end{figure}

In simulations, we set one diffusivity via $\tau = 1.0$ and $\theta = 0.5$, and set the other diffusivity $\tau$ value lower to achieve the proper diffusion constant for that ratio $R$, using Equation \ref{eqn:D}:
\begin{equation}
\tau = \frac{D}{\theta} + \frac{1}{2} = \frac{4D+1}{2}
\label{eqn:tau}
\end{equation}
As discussed previously, this minimizes numerical error. Figure \ref{fig:step} shows a diagram of such a step function.

Unlike in the constant diffusivity case, we now have two regimes of interest: dry and wet. It is instructive to consider the expected behavior of extreme values for the step function $D(\rho)$. First, in the asymptotic (but nonphysical) case where $\rho_c \to 0$, the polymer is always in the wet state, and exhibits the same behavior as in the case of constant diffusivity; the long-time periodic substrate concentration must therefore scale with the cycle timings. Similarly, if $\rho_c \to \rho_0 = 1$, the polymer is always in the dry state, with the same result. The only difference between the two scenarios lies in the constant diffusivity value in the regime of interest, which reduces to a scaling problem corresponding to a different value of the dimensionless constant $F$, and therefore provides no new information about the system.

\begin{figure}
\centering
\includegraphics[width=\columnwidth,clip=true]{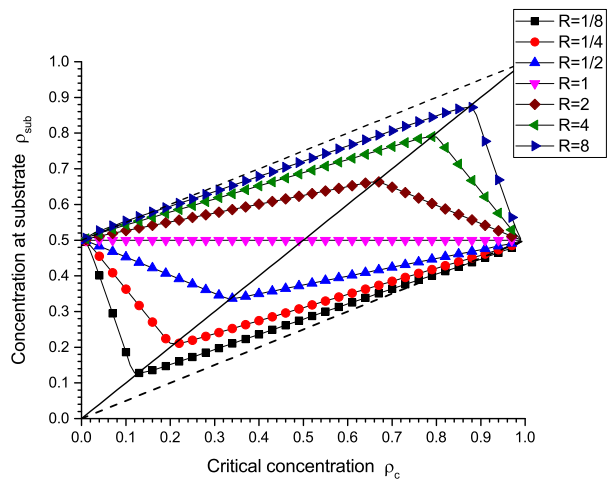}
\caption{Concentration at substrate $\rho_{\operatorname{sub}}$ versus step function critical concentration $\rho_c$ for selected diffusivity ratios $R$; all simulations use 1:1 cycle timing ratio. Peaks occur on the solid line $\rho_{\operatorname{sub}}=\rho_c$, and dashed lines represent the asymptotic values for $R \to \infty$ and $R \to 0$}
\label{fig:1-1-cycle}
\end{figure}

In the range $0 < \rho_c < 1$, the behavior is less obvious. One expects a higher periodic substrate concentration in this range for $R>1$ due to the presence of variable timescales, but it is not immediately clear what form the effect should take. We examine a range of diffusivity ratios $R$ numerically. For each, we ran a series of lattice Boltzmann simulations that vary the critical concentration $0 < \rho_c < 1$ and determine the substrate concentration oscillatory behavior at long times for cycle wet-to-dry ratios of 1:1, 1:2, 1:3, and 1:4, with the cycle timescale set by the fixed lower diffusivity. Since oscillations were found to be sufficiently small for the constant diffusivity case with $\tau = 1.0$, scaling to a lower value of $\tau$ does not pose any problems. While we ran the simulations for many diffusivity ratios, we plot the time-averaged steady-state results for only a representative few in Figure \ref{fig:1-1-cycle} for the 1:1 cycle ratio to illustrate the behavior.

The results for $\rho_c \to 0$ and $\rho_c \to \rho_0 = 1$ are clear; when the coating is either always dry or always wet, the resulting concentration scales precisely with the cycle timings regardless of the actual value of the diffusion constant used. Interestingly, the intermediate behavior shows two distinct linear regimes. For any given diffusivity ratio (\textit{i.e.} on one of the curves in Figure \ref{fig:1-1-cycle}), there is a maximal time-averaged substrate wetting for $R>1$, and a minimal wetting for $R<1$. Further, these extreme values lie on the solid line $\rho_{\operatorname{sub}} = \rho_c$ shown in the figure, implying that this occurs precisely at the corresponding swelling critical concentration.

An interesting asymptotic case occurs when either $R \to \infty$ or $R \to 0$. This corresponds to the coating having extremely low (resp. high) diffusivity before reaching the critical concentration; that is, when $D_{\dry} \to 0$ (resp. $D_{\wet} \to 0$). Effectively, this is equivalent to a scaled system whose concentration is allowed to vary only in the range $\rho_c \leq \rho \leq \rho_0 = 1$ for $R \to \infty$, or in the range $0 \leq \rho \leq \rho_c$ for $R \to 0$. In such a system, the eventual periodic concentration at the substrate must follow the dashed lines $\rho_{\operatorname{sub}} = \frac{\rho_0}{2}(1+\rho_c)$ for $R \to \infty$, or $\rho_{\operatorname{sub}} = \frac{\rho_0}{2}\rho_c$ for $R \to 0$, shown in Figure \ref{fig:1-1-cycle}.

\begin{figure}
\centering
\includegraphics[width=\columnwidth,clip=true]{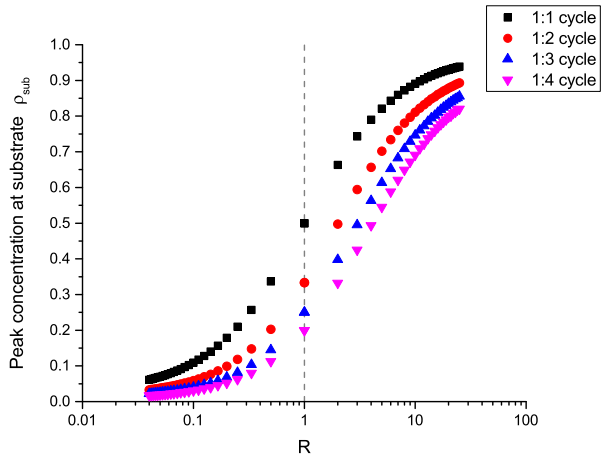}
\caption{Step function model extreme concentration at substrate $\rho_{\operatorname{sub}}$ versus diffusivity ratio $R$, given for multiple cycle timing ratios; values at $R=1$ (dashed) follow cycle timing ratio}
\label{fig:step-avg}
\end{figure}

What is unclear from first principles, however, is how the extreme substrate concentration value varies with either the diffusivity ratio or the cycle timing ratio. Figure \ref{fig:step-avg} plots the location of these extrema as a function of diffusivity ratio.

The values for $R=1$, corresponding to the constant-diffusivity case, exhibit the cycle timing ratio scaling discussed earlier. When the coating is allowed to dry for increasing time intervals relative to wetting, the extreme substrate wetting is reduced for all $R$; however, this reduction is far less effective farther from $R=1$. In any case, both the step function magnitude and cycle time ratio play a large role in the ``worst-case" substrate wetting that can arise for larger diffusivity ratios. We discuss the range $R<1$ in Section \ref{sec:material}.

\subsection{Variable diffusivity: linear}
While a step function represents a simple and useful model for concentration-dependent diffusivity arising from polymer network swelling, it is not the only such model. Most polymer networks exhibit a more gradual swelling behavior, making a step function only an approximation to physical behavior.

We consider here the effect of a linear change to diffusivity. In this model, the completely dry coating permits the slowest (but nonzero) moisture transport rate, which increases linearly to the fastest rate when fully saturated:
\begin{eqnarray}
D = D(\rho) &=& (D_{\wet}-D_{\dry})\rho + D_{\dry} \\
&=& D_{\dry}\left[(R-1)\rho + 1\right]
\end{eqnarray}
Materials for which $R<1$ result in a negative slope, discussed further in Section \ref{sec:material}. Analysis of this linear model is in some sense a simpler process, since it reduces to a single parameter $R$.

\begin{figure}
\centering
\includegraphics[width=\columnwidth,clip=true]{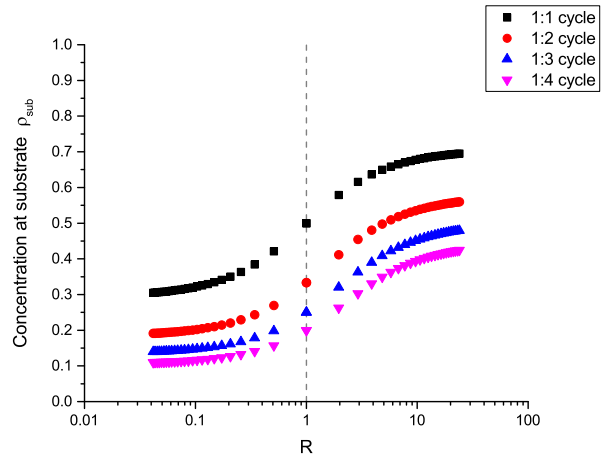}
\caption{Linear model concentration at substrate $\rho_{\operatorname{sub}}$ versus diffusivity ratio $R$, given for multiple cycle timing ratios; values at $R=1$ (dashed) follow cycle timing ratio}
\label{fig:linear-avg}
\end{figure}

Similarly to the step function case, we run a series of lattice Boltzmann simulations that fix the dry diffusivity using Equation \ref{eqn:tau} and vary the saturated diffusivity (effectively setting the slope of the linear dependence). The simulations run cycle time ratios of 1:1, 1:2, 1:3, and 1:4, with the overall cycle time fixed and the timescale set by the average of the two diffusivity values. We allow the system to reach long-time periodic behavior and examine the substrate time-averaged concentration. Figure \ref{fig:linear-avg} shows the time-averaged value of the substrate long-time oscillations for varying diffusivity and cycle time ratios.

Even though the diffusivity values at the dry and saturated extremes are chosen to be the same for both the step function and linear models for any given ratio $R$, the behavior is markedly different. For any given diffusivity ratio, the resulting concentration values in the linear model are more closely clustered near the $R=1$ value than their step function extreme value counterparts.

\section{Application to optimal material properties}
\label{sec:material}
We have so far discussed the effects of a diffusivity ratio $R>1$; that is, when a swollen coating permits faster moisture transport, either by an instantaneous increase in diffusivity at a critical concentration or more gradually in a linear manner. However, it is insightful to consider the symmetric case when $R<1$. This corresponds in the step function model to reversing the roles of the wet and dry constant values, and in the linear model to a negative slope. This would represent the behavior of a material that inhibits moisture transport at higher concentration levels.

In Figure \ref{fig:1-1-cycle}, bilinear curves for $R<1$ lie below the line $\rho_{\operatorname{sub}}=0.5$, and by symmetry each exhibits a distinct minimum value that lies on the line $\rho_{\operatorname{sub}}=\rho_c$ and decreases as $R \to 0$. Hence, for any given diffusivity ratio, there is a choice of critical concentration that minimizes the long-time saturation of the substrate below the value dictated by the cycle time ratio. This is in sharp contrast to materials for which $R>1$, where the cycle time values represent the ``best" case for wetting, and any other critical concentration results in greater substrate wetting over time. Of course, any cycle time ratio that increases the relative dry time will also reduce the overall substrate wetting, even for $R<1$ materials.

We see the same effect in the linear $R<1$ case in Figure \ref{fig:linear-avg}. There, the model provides no critical concentration to vary, and the mechanics of diffusion are set solely by the slope of $D(\rho)$. Any choice of $R<1$ results in eventual substrate saturation lower than otherwise dictated by the cycle time ratio, and is further affected by the relative wet and dry times of that cycle structure.

These results have implications for possible research into optimal single-layer materials for inhibiting corrosion. A metamaterial designed to slow diffusive processes with concentration leads to a far lower long-time substrate saturation than would otherwise be possible for an idealized material that allowed constant, and even extremely low, diffusivity.

\section{Summary and conclusions}
A discrete lattice Boltzmann method was used to model a finite coating test system consisting of an infinite moisture reservoir, a finite idealized barrier coating, and an impermeable substrate. The reservoir may be set to any concentration at any time. This models both natural environmental exposure and the common scenario when a coating is prepared on a test panel and placed into an environmental chamber for cyclic testing, where it is exposed to different moisture levels for long periods of time in order to determine its robustness for later use in service. An analytical solution for the concentration over time was presented, allowing for arbitrary cycling under the assumption of constant diffusivity. The numerical simulations matched the analytical solution with excellent accuracy.

We used the lattice Boltzmann simulations to determine the effects of moisture cycling at different cycle time ratios on the oscillatory concentration $\rho$ at the substrate after long times. Since the onset of corrosion of a panel system is commonly linked to exposure of the substrate to moisture, it is important to understand how $\rho$ evolves and stabilizes under different exposure regimes.

In the case where diffusivity is constant with concentration (as in the idealized Fickian case often used for barrier coatings), the steady-state behavior scales as expected with the cycle time ratio due to simple time averaging. However, many types of polymeric materials swell in the presence of sufficient solvent, increasing the rate of diffusivity. We considered two simple forms for the concentration-dependent diffusivity $D(\rho)$: a step function, where the polymer network is collapsed until a critical concentration is reached and swollen thereafter, and a linear model, where the dry network has a nominal diffusivity that increases until it reaches a maximal level when the coating is fully saturated.

Under the step function model, all parameters in the model affect the long-time oscillatory concentration. Regardless of the step function parameters (critical concentration and diffusivity change after wetting), increasing the relative dry time in a cycle protocol leads to a lower steady-state substrate concentration. As the critical concentration varies from very low (where the coating is almost always in the wet state) to very high (where it is almost always in the dry state), the long-time value increases linearly, reaches a peak whose location is fixed by the wet diffusivity, and thereafter decreases linearly.

Using the linear model, the behavior is similar, but much different in extent. As in the step function model, the cycle timing ratio has an effect on the overall behavior, but it is minimal compared to the overall saturation.

In either model, the results indicate that there is a second very different, but symmetric, regime that inverts the wet and dry diffusivity values. For traditional materials that permit faster transport under saturation, the choice of material properties can minimize substrate wetting to a value that will be greater than the constant diffusivity case for any cycle protocol. However, a metamaterial structurally designed to inhibit transport under saturation would permit an optimization of properties allowing for substrate wetting far lower than under constant diffusivity, even under cycle protocols with high relative wet periods.

These results imply that moisture cycling can play a large role in understanding the timescales and concentrations that may lead to corrosion, especially when considering the types of environmental tests that are regularly performed on candidate coating systems. Whether or not a particular type of polymeric system is subject to swelling, as in the case of a good urethane barrier topcoat versus an epoxy primer, plays a much larger role in the long-time behavior, and the parameters and form of the swelling lead to large variations. Since electrochemical means of determining diffusion and uptake behavior, such as impedance spectroscopy, rely on constant electrolyte exposure so that a useful equivalent circuit is applicable, controllable simulations such as these offer unique insight into variations caused by moisture cycling, and offer a path toward the design of better coating systems and test protocols.

\section{Acknowledgments}
The authors thank Kyle Strand of North Dakota State University for ongoing discussions and helpful insights. Computational support from the North Dakota State University Center for Computationally Assisted Science and Technology is gratefully acknowledged. The first author was supported by the Strategic Environmental Research and Development Program under contract W912HQ-15-C-0012. Views, opinions, and/or findings contained in this report are those of the authors and should not be construed as an official Department of Defense position or decision unless so designated by other official documentation.

\bibliography{main}

\end{document}